\documentclass{ws-p10x7}

\begin{document}

\title{Signature neutrinos from ultrahigh-energy photons}


\author{Alexander Kusenko}

\address{Department of Physics and Astronomy, UCLA, Los Angeles, CA
90095-1547 \\ and \\ RIKEN BNL Research Center, Brookhaven National
Laboratory, Upton, NY 11973 
\\E-mail: kusenko@ucla.edu}

\twocolumn[\maketitle\abstract{ At high red shift, the temperature of
cosmic microwave background is sufficiently high for the ultrahigh-energy
photons to pair-produce muons and pions through interactions with the
background photons.  At the same time, the radio background and magnetic
fields are too weak to drain energy out of the electromagnetic cascade
before the muons and pions are produced.  Decays of the energetic muons and
pions yield neutrinos with some distinctive spectral properties that can be
detected and can indicate the presence of ultrahigh-energy photons at high
red shift.  The neutrino signature can help identify the origin of cosmic
rays beyond the Greisen-Zatsepin-Kuzmin cutoff.  }]

The origin of ultrahigh-energy cosmic rays~\cite{data}, with energies beyond
the Greisen-Zatsepin-Kuzmin (GZK) cutoff~\cite{gzk}, remains an outstanding
puzzle~\cite{reviews}.  Many proposed explanations invoke new sources,
such as superheavy relic particles~\cite{particles,kt} or topological
defects~\cite{TD_reviews,TD}, which can generate photons at both low
and high red shifts.  To understand the origin of the ultrahigh-energy
cosmic rays (UHECR), it is crucial to distinguish such sources from more
conventional astrophysical ones~\cite{conv}.  The latter tend to produce
more protons than photons.  In addition, the ``astrophysical'' candidate
sources like, {\em e. g.}, active galactic nuclei,  have formed at
relatively low red shift.  In contrast, topological defects could operate
at much higher red shifts. 

Sources of ultrahigh-energy photons that were active at red shift $z>3$ can
be identified by observation of neutrinos produced in interactions of
energetic and background photons~\cite{kp}.  This may help understand the
origin of UHECR.

At red shift $z$ the cosmic microwave background radiation (CMBR) has
temperature $T_{_{CMB}}(z) = 2.7 (1+z){\rm K}$.  Because of this, at high
red shift the photon-photon interactions can produce pairs of muons and
charged pions, whose decays generate neutrinos. This is in sharp contrast
with the $z \stackrel{<}{_{\scriptstyle \sim}} 1$ case, where the photons
do not produce neutrinos\footnote{Neutrinos can be
produced by sources of ultrahigh-energy protons through pion
photoproduction~\cite{nus}. Here we only consider sources of
ultrahigh-energy photons.} as they lose energy mainly by scattering off the
radio background through electron-positron pair production and subsequent
electromagnetic cascade~\cite{berezinsky}.  The ratio of the CMBR density
to that of universal radio background (RB) increases at higher $z$, and the
process $\gamma \gamma_{_{CMB}}\rightarrow \mu^+ \mu^- \rightarrow e^+e^-
\bar{\nu}_\mu \nu_\mu \bar{\nu}_e \nu_e $ can produce neutrinos.  The
threshold for this lowest-energy neutrino-generating interaction is
$\sqrt{s}> 2 m_\mu=0.21$GeV, or
\begin{equation}
E_\gamma > E_{\rm th}(z)=\frac{ 10^{20}{\rm eV}}{1+z}
\label{threshold}
\end{equation}

At $z<1$ the main source of energy loss for photons is electromagnetic
cascade that involves $e^+e^-$ pair production (PP) on the radio background
photons~\cite{berezinsky,reviews}.  The radio background is generated by
normal and radio galaxies.  Its present density~\cite{biermann} is higher
than that of CMB photons in the same energy range.  The radio background
determines the mean interaction length for the $e^+e^-$ pair production.
At red shift $z$, however, the comoving density of CMB photons is the same,
while the comoving density of radio background is lower.  Models of
cosmological evolution of radio sources~\cite{condon} predict a sharp drop
in the density of radio background at red shift $z
\stackrel{>}{_{\scriptstyle \sim}} 2$.  Let $z_{_R}$ be the value of red
shift at which the scattering of high-energy photons off CMBR dominates
over their scattering off RB.  Based on the models of RB~\cite{condon}, we
take $z_{_R} \sim 3$.  Another source of energy losses in the
electromagnetic cascade is the synchrotron radiation by the electrons in
the intergalactic magnetic field (IGMF)~\footnote{I thank V.~Berezinsky for
pointing this out to me.}.  This is an important effect for red shift
$z<z_{_M}$, where $z_{_M} \sim 5$ corresponds to the time when IGMF is
weak, and the synchrotron losses are not significant.  

Let us now consider the propagation of photons at $z>z_{\rm min}=\max
(z_{_R},z_{_M})$.  In 
particular, we are interested in the neutrino-generating process $\gamma
\gamma_{_{CMB}} \rightarrow \mu^+ \mu^-$. The threshold for this reaction
is given in eq.~(\ref{threshold}).  For $\sqrt{s}> 2m_{\pi^{\pm}} =
0.28$~GeV the charged pion production and decay can also contribute to the
neutrino flux.
Although the cross section for the electron pair production is higher than
that for the muon pair production,  neutrinos are nevertheless
produced.  This is because the high-energy photons are continuously
regenerated in the electromagnetic cascade~\cite{reviews}.  
Since the energies of the two interacting photons are vastly different,
either the electron or the positron produced in the reaction $\gamma
\gamma_{_{CMB}}\rightarrow e^+e^-$ has energy close to that of the initial
photon. This electron undergoes inverse Compton scattering (ICS) and
produces a photon with a comparable energy.  As a result, the
electromagnetic cascade creates a mixed beam of photons and electrons with
comparable fluxes.  Thanks to the regeneration of high-energy photons, the
energy attenuation length $\lambda_{\rm eff}$ is much greater than the pair
production interaction length $\lambda(\gamma \gamma_{_{CMB}}\rightarrow
e^+e^-)$.  

For energies in the range of interest, $\lambda_{\rm eff} \gg \lambda(\gamma
\gamma_{_{CMB}}\rightarrow \mu^+\mu^-)$.  Therefore, in the absence of
dense radio background all photons with $E>  E_{\rm th}$ pair-produce muons
and pions before their energy is reduced by the cascade.  
Due to the kinematics, one of the two muons has a much higher
energy than the other, in full analogy with the $e^+e^-$ case.  Muons decay
before they can interact with the photon background.  Each energetic muon
produces two neutrinos and an electron.  The latter can regenerate a photon
via ICS. This process can repeat until the energy of a regenerated photon
decreases below the threshold for muon pair production.  

The flux of neutrinos relative to the observed flux of UHECR depends on the
time evolution of the source.  Sources of UHE photons, whether they are
topological defects~\cite{TD} or decaying relic particles with cosmologically
long lifetimes~\cite{particles}, produce  high-energy photons at some
rate $\dot{n}_{_X}$.  One can parameterize~\cite{TD} this rate as
$\dot{n}_{_X} =\dot{n}_{_X,0} (t/t_0)^{-m}$, with $m=0$ for decaying relic
particles, $m=3$ for ordinary string and necklaces, and $m
\ge 4$ for superconducting strings~\cite{TD_reviews,TD}.  

Sources with $m=3$ are of particular interest as possible candidates for
the origin of UHECR~\cite{TD_reviews,TD}.  The neutrinos produced by such
sources at high red shift have energies~\cite{kp} up to $10^{18}$~eV with a
sharp cutoff below $10^{19}$eV.  The flux of $10^{18}$~eV
neutrinos~\cite{kp} is predicted to be $ \sim 10^{-16}{\rm cm}^{-2} {\rm
s}^{-1} {\rm sr}^{-1}$.  This flux will be accessible to several experiments
in the near future~\cite{exp}.

In summary, sources of ultrahigh-energy photons that operate at red shift
$z > z_{\min}\sim 5$ produce neutrinos with energy
$E_\nu \sim 10^{18}$eV.  The flux depends on the evolution index $m$ of the
source.  A distinctive characteristic of this type of neutrino background
is a cutoff below $10^{19}$eV due to the universal radio background and
magnetic fields at $z<z_{\rm min}$.  This is in contrast with sources of
ultrahigh-energy protons that can produce neutrinos with energies up to the
GZK cutoff and beyond.

I thank V.~Berezinsky for very helpful discussions.  
This work was supported in part by the US Department of Energy grant
DE-FG03-91ER40662, Task C, as well as by a 
grant from UCLA Council on Research.


\end{document}